\documentclass[12pt,oneside, a4paper]{article}

\pdfoutput=1  
 
\ifx\pdfoutput\undefined
\usepackage[dvips,bookmarks=false]{hyperref}	
\else
\usepackage{hyperref}	
\fi
\hypersetup{colorlinks,bookmarksopen,bookmarksnumbered,citecolor=blue,
linkcolor=black,pdfstartview=FitH,urlcolor=blue}

\usepackage{graphicx}
\usepackage{subfigure}
\usepackage{amssymb}
\usepackage{cite}
\usepackage{bm}
\usepackage{amsmath,amsthm}

\numberwithin{equation}{section}

\newcommand{\capdef}{}
\newcommand{\mycaption}[2][\capdef]{\renewcommand{\capdef}{#2}%
       \caption[#1]{{\footnotesize #2}}}

\newcommand{\be}{\begin{equation}}
\newcommand{\ee}{\end{equation}}

\newcommand{\Sum}{\ensuremath{\Sigma}}

\begin{document}

\begin{titlepage}

\begin{center}

\vspace*{2cm}
{\Large\bf Cosmology and the neutrino mass ordering}
\vspace{1cm}

\renewcommand{\thefootnote}{\fnsymbol{footnote}}
{\bf Steen Hannestad}$^a$\footnote[1]{sth@phys.au.dk} 
				and  
{\bf Thomas Schwetz}$^b$\footnote[2]{schwetz@kit.edu}
\vspace{5mm}

{\it%
$^a${Department of Physics and Astronomy, Aarhus University,\\ Ny Munkegade, DK-8000 Aarhus C, Denmark} \\  
$^b${Institut f\"ur Kernphysik, Karlsruhe Institute of Technology (KIT),\\ 76021 Karlsruhe, Germany}
}

\vspace{8mm} 

\abstract{We propose a simple method to quantify a possible exclusion
  of the inverted neutrino mass ordering from cosmological bounds on
  the sum of the neutrino masses. The method is based on Bayesian
  inference and allows for a calculation of the posterior odds of normal
  versus inverted ordering. We apply the method for a specific set of
  current data from Planck CMB data and large-scale structure surveys,
  providing an upper bound on the sum of neutrino masses of 0.14~eV at
  95\%~CL. With this analysis we obtain posterior odds for normal
  versus inverted ordering of about 2:1. If cosmological data is
  combined with data from oscillation experiments the odds reduce to
  about 3:2. For an exclusion of the inverted ordering from cosmology at
  more than 95\%~CL, an accuracy of better than 0.02~eV is needed for
  the sum.  We demonstrate that such a value could be reached with
  planned observations of large scale structure by analysing
  artificial mock data for a EUCLID-like survey.  }

\end{center}
\end{titlepage}

\renewcommand{\thefootnote}{\arabic{footnote}}
\setcounter{footnote}{0}

\setcounter{page}{2}

\section{Introduction}

Current data on neutrino oscillations show a degeneracy between two
possible orderings of the neutrino mass states, the normal ordering
(NO) and inverted ordering (IO). Breaking this degeneracy is one of
the main goals of upcoming oscillation experiments,
e.g.,~\cite{Messier:2013sfa, Acciarri:2015uup, Aartsen:2014oha,
  Djurcic:2015vqa, Ahmed:2015jtv}, see \cite{Blennow:2013oma} for an
overview. On the other hand, also cosmological observations
potentially may contribute to this question. Cosmological structure
formation is sensitive mostly to the sum of the neutrino masses,
$\Sum$.  There are subtle effects sensitive to the details of the
neutrino mass spectrum beyond the sum, see e.g.,
\cite{Lesgourgues:2004ps, Slosar:2006xb, Jimenez:2010ev,
  Hamann:2012fe}.  With realistic observations in the foreseeable
future those effects will be very hard to detect \cite{Hamann:2012fe}.
Focusing on the sum of masses, we can use that oscillation data determine the
mass-squared differences and we have:
\begin{align}\label{eq:sum}
  \Sum  \equiv \sum_{i=1}^3m_i = \left\{ 
  \begin{array}{l@{\qquad}l}
  m_0 + \sqrt{\Delta m^2_{21} + m_0^2} + \sqrt{\Delta m^2_{31} + m_0^2} & \text{(NO)} \\    
  m_0 + \sqrt{|\Delta m^2_{32}| + m_0^2} + \sqrt{|\Delta m^2_{32}| - \Delta m^2_{21} + m_0^2} &
  \text{(IO)} \\    
  \end{array} \right. \,,
\end{align}
where $m_0$ denotes the lightest neutrino mass, where by convention $m_0 \equiv m_1 \,
(m_3)$ for NO (IO). The mass-squared differences $\Delta m^2_{ij} \equiv m_i^2 - m_j^2$ are
determined to \cite{Gonzalez-Garcia:2014bfa} ($1\sigma$ uncertainties):
\begin{align} \label{eq:dmq}
  \Delta m^2_{21} = 7.49^{+0.19}_{-0.17} \times 10^{-5} \, \text{eV}^2 \,,\qquad
  \begin{array}{l@{\quad}l}
    \Delta m^2_{31} = \,\,\, 2.484^{+0.045}_{-0.048} \times 10^{-3}\, \text{eV}^2 & \text{(NO)} \\ 
    \Delta m^2_{32} = -2.467^{+0.041}_{-0.042} \times 10^{-3}\, \text{eV}^2 & \text{(IO)}  
  \end{array} \,.
\end{align}

For a zero lightest neutrino mass ($m_0 = 0$), the predictions for the sum are
($1\sigma$ uncertainties)
\begin{align}
  \Sum = \left\{ 
  \begin{array}{l@{\qquad}l}
   58.5 \pm 0.48 \, \text{meV} & \text{(NO)} \\    
   98.6 \pm 0.85 \, \text{meV} & \text{(IO)} \\    
  \end{array} \right. \qquad (m_0 = 0)\,.
  \label{eq:min-values}
\end{align}
Hence, if cosmological observations provide a determination of
\Sum\ significantly below 0.098~eV, the inverted mass ordering would be
disfavoured.

Recent data from Planck CMB data combined with baryonic acoustic
oscillations (BAO) and other observations lead to the bound $\Sum <
0.23$~eV at 95\%~CL (PlanckTT + lowP + lensing + BAO + JLA + $H_0$),
see \cite{Ade:2015xua} for details. Depending on the used data and
variations in the analysis, different authors obtain upper bounds from
current data approaching the ``critical'' value of 0.1~eV
\cite{Palanque-Delabrouille:2015pga,Cuesta:2015iho, Huang:2015wrx, DiValentino:2015sam,
  Giusarma:2016phn}. These results suggest that IO starts to get under pressure from cosmology.

In this note we want to point out that such a claim should be based on
a proper statistical analysis. The question to be answered is, whether
the hypothesis of IO can be rejected with some confidence against NO.
For a related discussion in the context of oscillation experiments see
for instance ref.~\cite{Blennow:2013oma} formulated in terms of
frequentist hypothesis testing, or ref.~\cite{Blennow:2013kga} using
Bayesian reasoning. Indeed, just from the numbers in
eq.~\eqref{eq:min-values} one sees that it is not enough that the
upper bound on $\Sum$ is below 0.098~eV, but instead cosmology needs to
determine \Sum\ with an accuracy better than about 0.02~eV in order to
exclude a value of 0.098~eV against 0.059~eV at $2\sigma$. Note that
this would imply a $ \gtrsim 3\sigma$ detection of a non-zero value of
$\Sum \approx 0.058$~eV. Obviously, requirements would be even more
demanding if $m_0$ turns out not to be zero. Below we are going to
substantiate this simple estimate by more detailed calculations.

\section{Quantifying the evidence against inverted ordering}
\label{sec:method}

In this section we provide a simple recipe to quantify possible
evidence against inverted ordering from cosmology. Note that as long
as only \Sum\ is the dominating observable, it will never be possible
to reject NO. We will use Bayesian statistics, following closely
\cite{Blennow:2013kga}. Similar methods have been used in
  \cite{Hall:2012kg} in the context of the mass ordering in
  cosmology. Bayesian methods are especially suitable for our
problem, since we are interested in a region close to a physical
boundary implied by $m_0 \ge 0$. Indeed, the mechanism to exclude IO
is based on the fact that the data may prefer a value of $\Sum$
outside the physical domain accessible in the case of IO. Such a
situation is easily incorporated in Bayesian statistics. In a
frequentist approach, the relevant distribution of a test statistics
needs to be obtained by Monte Carlo simulations, since one expects
non-Gaussian behaviour close to a physical boundary.

We consider the likelihood function $\mathcal{L}(D|\theta,m_0,O)$, of
some set of cosmological data $D$, with the theoretical model
depending on a set of cosmological parameters $\theta$, the lightest
neutrino mass, $m_0$, and the discrete parameter $O$ describing
the mass ordering, $O = N,I$. Using Bayes theorem, we easily obtain
the probability for a mass ordering given data $D$:
\begin{align}
  p_O \equiv p(O|D) = \frac{\pi(O)}{\pi(D)}  \int d\theta \int dm_0 \,
  \mathcal{L}(D|\theta,m_0,O) \pi(\theta) \pi(m_0) \,,
\end{align}
where the $\pi$ denote prior probabilities. Defining the likelihood
marginalized over cosmological parameters as $\mathcal{L}(D|m_0,O)
\equiv \int d\theta \, \mathcal{L}(D|\theta,m_0,O) \pi(\theta)$ and
adopting a flat prior for $m_0 \ge 0$ we obtain
\begin{align}\label{eq:prob-MO}
  p_O = \frac{ \pi(O) \int_0^\infty dm_0 \, \mathcal{L}(D|m_0,O)}
  {\pi(N) \int_0^\infty dm_0 \, \mathcal{L}(D|m_0,N) +
    \pi(I) \int_0^\infty dm_0 \, \mathcal{L}(D|m_0,I)}
\end{align}
with $p_N + p_I = 1$. If no prior information on the mass ordering is
available an obvious choice is to assume that NO and IO are equally
likely a priori: $\pi(N) = \pi(I) = 1/2$. However, using $\pi(O)$ it
is straight forward to include possible prior information on the
ordering from oscillation data. The Bayesian analysis of
\cite{Bergstrom:2015rba} gives for present oscillation data a
posterior probability for IO of 0.55 (i.e., very close to equal
probabilities for NO and IO). However, this may improve in the near
future by upcoming oscillation data.

Using eq.~\eqref{eq:prob-MO}, one can then consider for instance the
ratio $p_I/p_N$ to define the posterior odds of IO versus NO
\cite{Blennow:2013kga}.  Alternatively one can report $p_I$ to
quantify how likely an inverted mass ordering is for given
data. Values of $p_I \ll 1$ will provide exclusion of IO at a
confidence level of $(1-p_I)$.

\section{Analysis of cosmological data}
\label{sec:data}

\subsection{Current data}

In any parameter estimation analysis of cosmological data a model has
to be specified. A larger parameter space inevitably leads to less
stringent bounds on the neutrino mass (and other cosmological
parameters). In the standard $\Lambda$CDM model data from the Planck
mission provides an upper bound $\Sum < 0.72$~eV at 95\%~CL
(``PlanckTT + lowP''); with the addition of large scale structure data
this upper bound improves to $\Sum < 0.23$~eV (``PlanckTT + lowP +
lensing + BAO + JLA + $H_0$''), see \cite{Ade:2015xua} for details of
the used data.

However, the parameter space used in this model is quite restrictive
and fits data using only 6 parameters in addition to the sum of
neutrino masses: $\Omega_b h^2$, the physical baryon density,
$\Omega_c h^2$, the physical cold dark density, $H_0$, the Hubble
parameter, $A_s$, the amplitude of the primordial scalar fluctuation
spectrum, $n_s$, the spectral tilt of the primordial spectrum, and
$\tau$, the optical depth to reionization.  In more general parameter
spaces these bounds can be relaxed significantly (see
e.g.\ \cite{DiValentino:2016hlg,Lu:2016hsd,Canac:2016smv} for recent
examples of extended models). Since our goal here is not so much to
see to what extent the neutrino mass bound can be relaxed, but rather
a study of the sensitivity to reject the IO already based on current
data we will use the restricted parameter space defined by the 6
parameter $\Lambda$CDM model with the addition of $\Sum$.

For our analysis we use the Planck 2015 data, including polarisation
\cite{Ade:2015xua}.  We furthermore include BAO data from a variety of
different surveys: 6dFGS~\cite{Beutler:2011hx},
SDSS-MGS~\cite{Ross:2014qpa}, BOSS-LOWZ \cite{Anderson:2012sa} and
CMASS-DR11~\cite{Anderson:2013zyy}.  Finally, we also include the
recent local universe measurement of the Hubble parameter, $H_0=73.02
\pm 1.79 \,\, {\rm km} \, {\rm s}^{-1} \, {\rm
  Mpc}^{-1}$~\cite{Riess:2016jrr}.\footnote{Within the minimal
  $\Lambda$CDM model this local value for $H_0$ is more than $3\sigma$
  away from the global result from Planck, see e.g.,
  \cite{DiValentino:2016hlg} for a discussion. We have checked that
  our constraints for \Sum\ are not sensitive to this tension in the
  data, and we obtain indistinguishable results for \Sum\ as well as
  for the probability of IO without using the local $H_0$ prior. We
  note that this would have been different in an extended analysis
  with more free parameters. In that case the $H_0$ measurement is
  important for breaking e.g.\ the degeneracy between \Sum\ and
  $N_{\rm eff}$.}  To perform parameter estimation and derive
constraints we have used the publicly available CosmoMC code
\cite{Lewis:2002ah}.

In this relatively restricted model we find an upper bound of $\Sum <
0.14$~eV (95\%~CL). This is comparable to other recent estimates using
somewhat different data sets and model assumptions
\cite{Palanque-Delabrouille:2015pga,Cuesta:2015iho, Huang:2015wrx,
  DiValentino:2015sam, Giusarma:2016phn}. Since our purpose here is
mainly to discuss what claims can be made about the neutrino mass
ordering given a constraint on $\Sum$ in this range we will not
explore how the bound changes with the use of different data and model
assumptions (this has been discussed in many other recent papers). In
order to check the approximation that current data is sensitive only
to the sum of neutrino masses we have performed three analyses with
the following assumptions: (i) two massless and one massive neutrino,
(ii) one massless and two degenerate massive neutrinos, and (iii)
three degenerate massive neutrinos. Note that none of these scenarios
actually corresponds to the realistic cases of NO or IO with
mass-squared differences constrained by oscillations. However, the
spread in the results will be indicative for our assumption that
cosmology is sensitive only to $\Sum$. Indeed we confirm that within
the numerical accuracy all three models lead to an upper bound of
0.14~eV (95\%~CL).

\begin{figure}[t]
  \centering
  \includegraphics[width=0.8\textwidth]{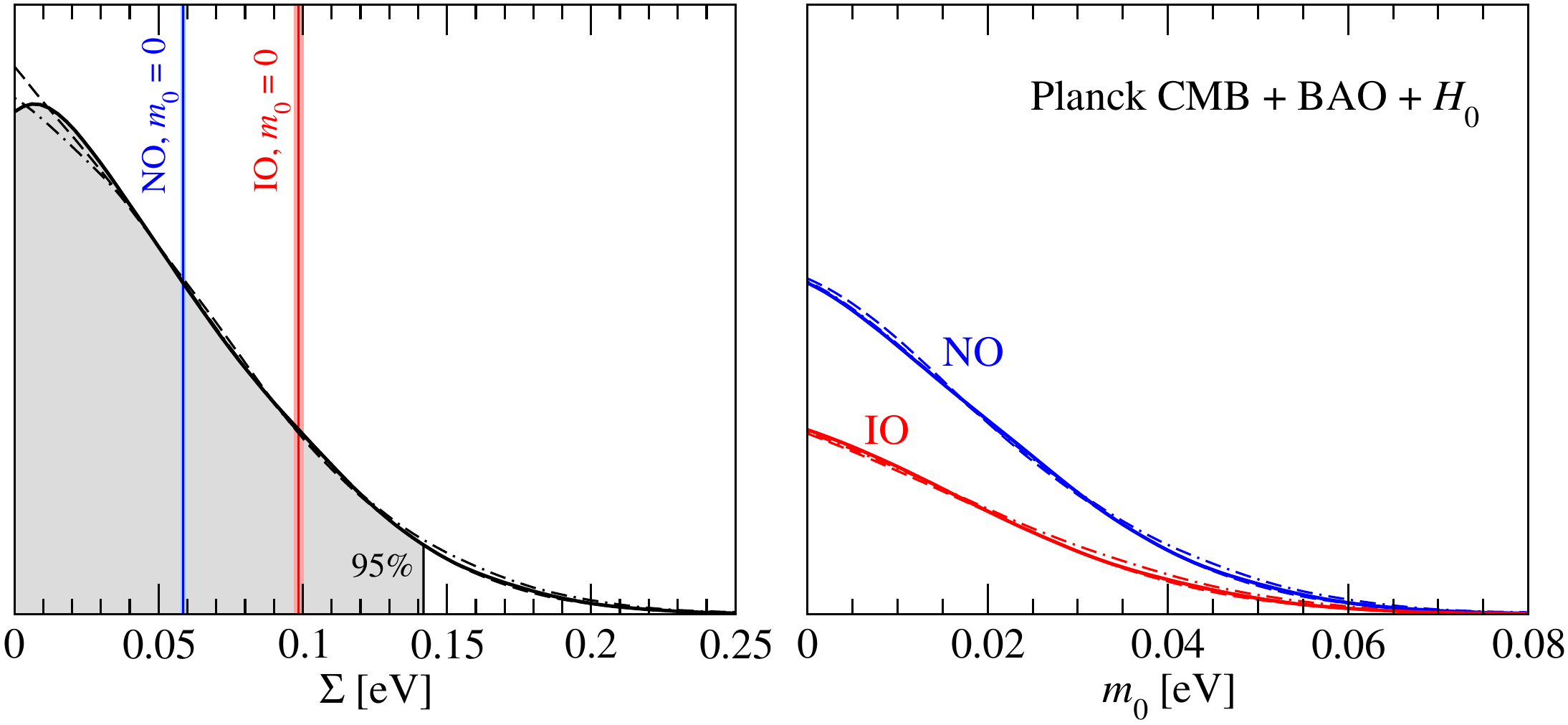}
  \mycaption{Posterior likelihood function from current data
    (Planck+BAO+$H_0$).  The left panel shows the posterior likelihood
    function for $\Sum$, where we indicate the predicted values for NO
    and IO in the case of $m_0 = 0$; the width of the lines
    corresponds to $\pm 2\sigma$ uncertainty due to current
    oscillation data. The gray shaded region indicates the one-sided
    upper bound on \Sum\ at 95\%~CL (flat prior in \Sum). The right
    panel shows the posterior likelihood as a function of $m_0$ for NO
    and IO with appropriate relative normalization. The dashed,
    dot-dashed, solid curves correspond to the approximation that 1,
    2, 3 massive neutrinos contribute to $\Sum$ (see text for
    details).
    \label{fig:LH-current}}
\end{figure}

The posterior likelihood function is shown in
fig.~\ref{fig:LH-current}.  The left panel shows the likelihood as a
function of \Sum, and we indicate the predicted values for $\Sum$ for
NO and IO assuming $m_0=0$, as well as the 95\%~CL upper bound on
\Sum, assuming a flat prior in $\Sum \ge 0$. Note that the region of
largest likelihood, for $\Sum < 59$~meV, is actually unphysical, since
such small values for the sum of the neutrino masses are inconsistent
with neutrino oscillation data. Hence, this region will be cut away
once the sum is expressed using eq.~\eqref{eq:sum} and imposing the
physical requirement of $m_0 \ge 0$.

In order to apply eq.~\eqref{eq:prob-MO} to calculate the probability
of IO vs NO we translate the likelihood into a posterior likelihood as
a function of $m_0$ by using eq.~\eqref{eq:sum}.\footnote{We neglect
  the uncertainty induced by the uncertainty on the mass-squared
  differences from oscillation data. For an accuracy on \Sum\ larger
  than 0.01~eV this is an excellent approximation, see also
  sec.~\ref{sec:gauss}.} The resulting likelihoods are shown in the
right panel of fig.~\ref{fig:LH-current}. The posterior odds for NO
versus IO are given by the ratio of the integrals over those two
curves weighted by the prior probabilities for the orderings. Assuming
equal prior probabilities for NO and IO, eq.~\eqref{eq:prob-MO} leads
to a probability for IO of $p_I = 0.35$, which corresponds to
posterior odds for NO versus IO of about 1.9:1. Clearly, using even
quite restrictive assumptions about the cosmological model current
data is not sufficient to distinguish between the NO and the IO at a
statistically significant level.

If instead of equal priors for the two orderings we use the prior
probabilities from oscillation data~\cite{Bergstrom:2015rba}, $\pi(I)
= 0.55, \, \pi(N) = 0.45$, we obtain a posterior probability of $p_I =
0.392$ or equivalently, posterior odds of 1.55:1 for NO vs IO. Again
this result shows that present data from neutrino oscillations and
cosmology are not sensitive enough to reach a significant
conclusion. However, this exercise does illustrate the power of the
method to combine information from oscillations and cosmology which is
expected to be very useful in the near future.

The different curves in fig.~\ref{fig:LH-current} (dashed, dot-dashed,
solid) correspond to the three different assumptions about how $\Sum$
is shared between the three neutrinos (1 massive, 2 massive, 3
massive, respectively). We see that the differences are small, and the
MO analysis gives identical results within the numerical
accuracy. This justifies our approximation that the likelihood depends
on $\Sum$ only when converting $\mathcal{L}(D|\Sum)$ into
$\mathcal{L}(D|m_0,O)$ by using eq.~\eqref{eq:sum}.

\subsection{Prospective data including a EUCLID-like survey}

Let us now address the question of how this situation will change
quantitatively in the future.  In the coming years a whole range of
new cosmological surveys will start operating, including for example
the LSST survey \cite{LSST} and the EUCLID satellite mission
\cite{euclid}.  When combined with CMB data these surveys have the
potential to bring the sensitivity to $\Sum$ down to the 0.02~eV level
(see e.g.\ \cite{Hamann:2012fe,Audren:2012vy,Basse:2013zua,Cerbolini:2013uya,Font-Ribera:2013rwa,Abazajian:2013oma, Allison:2015qca}). 

Using the CosmoMC-based forecasting tool described in \cite{Hamann:2012fe,Basse:2013zua,
Basse:2014qqa} we have generated artificial EUCLID-like data and used it to constrain 
cosmological parameters including $\Sum$.
Specifically we have used a EUCLID-like data set consisting of weak lensing and photometric galaxy 
survey components. We have used synthetic data equivalent to the ``csg'' case described in
\cite{Basse:2013zua}, which includes synthetic cosmic shear (s) and galaxy (g) data, as well as CMB data (c) roughly equivalent to Planck data in precision. As the cosmological model we have used a minimal $\Lambda$CDM model with $\Sum$ and the number of relativistic degrees of freedom, $N_{\rm eff}$, as additional parameters.

The fiducial model we use has one massive neutrino with $m_\nu = 0.06$ eV and 2.046 massless neutrinos. Note that this is not equivalent to the real physical prediction of the NO. However, from a cosmological parameter estimation point of view disentangling this model from the case with 1.015 neutrinos with mass 0.05 eV, 1.015 with mass 0.01 eV, and 1.015 massless requires much higher precision than what is projected for EUCLID \cite{Hamann:2012fe} (see e.g.\ \cite{Jimenez:2016ckl} for a recent treatment). Therefore this slightly simplified model is more than adequate for the purpose of this paper.

\begin{figure}[t]
  \centering \includegraphics[width=0.8\textwidth]{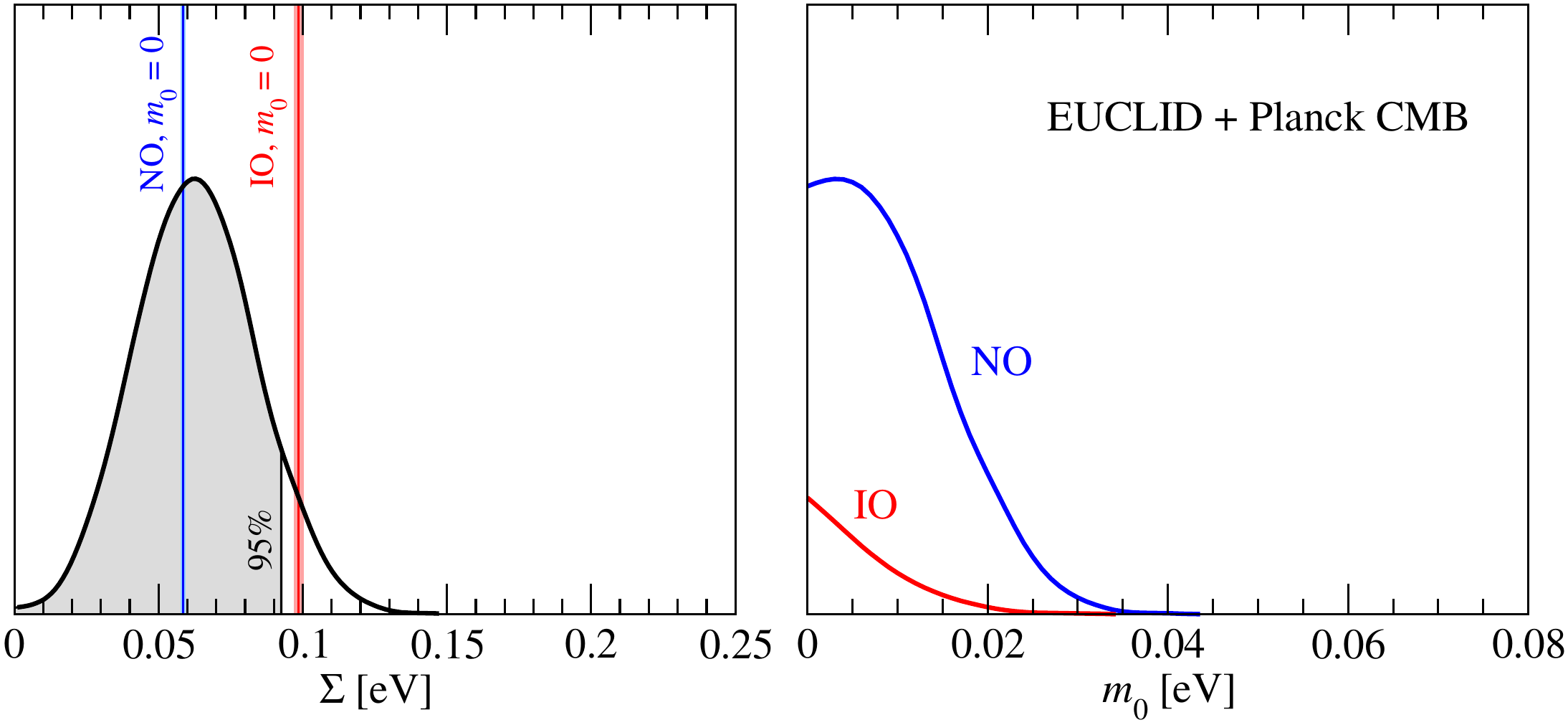}
  \mycaption{Posterior likelihood function from simulated future data
    (EUCLID+Planck CMB).  The left panel shows the posterior
    likelihood function for $\Sum$ for a fiducial model with one
    massive neutrino with $m_\nu = 0.06$~eV and two massless
    neutrinos. We indicate the predicted values for NO and IO in the
    case of $m_0 = 0$; the width of the lines corresponds to $\pm
    2\sigma$ uncertainty due to current oscillation data. The gray
    shaded region indicates the one-sided upper bound on \Sum\ at
    95\%~CL (flat prior in \Sum). The right panel shows the posterior
    likelihood as a function of $m_0$ for NO and IO with appropriate
    relative normalization.
    \label{fig:LH-future}}
\end{figure}

The resulting posterior likelihood as a function of \Sum\ is shown in
the left panel of fig.~\ref{fig:LH-future} and we obtain the formal
parameter constraint $\Sum = 0.060 \pm 0.021$~eV~(68\%
CL). Note that here we marginalize also over $N_{\rm eff}$
  in addition to the 6 parameters of the $\Lambda$CDM model. Similar
as above we transform the likelihood now into a likelihood for $m_0$
assuming either NO or IO, see right panel. We ignore
  the small effects of the different orderings of the neutrino masses
  and use the same likelihood to describe both normal and inverted
  orderings. As mentioned above this should be an excellent
  approximation for the used data set. The relative posterior
likelihood for NO and IO is given by the ratio of the areas under the
two curves. Assuming equal prior probabilities for NO and IO we
obtain a probability for IO according to eq.~\eqref{eq:prob-MO} of
8\%, which corresponds to posterior odds of NO versus IO of
approximately 12:1.

\section{Sensitivity estimates with a Gaussian toy likelihood}
\label{sec:gauss}

From fig.~\ref{fig:LH-future} one can see that the likelihood function as a function of \Sum\ is close to Gaussian. This is certainly true for the simulated EUCLID data, but holds approximately also for present data. To estimate the required accuracy needed on $\Sum$ to exclude IO we
assume therefore that the likelihood function from cosmology can be approximated by 
\begin{equation}
  \mathcal{L}(\Sum^{\rm obs} | m_0, O) = \frac{1}{\sqrt{2\pi} \sigma}
  \exp\left[ - \frac{(\Sum^{\rm obs} - \Sum(m_0,O) )^2}{2\sigma^2} \right]  
\end{equation}
where $\Sum(m_0,O)$ is given in eq.~\eqref{eq:sum}, and $\sigma^2 =
\sigma_{\rm osc}^2 + \sigma_{\rm obs}^2$, with $\sigma_{\rm osc}(m_0,
O)$ being the error on $\Sum$ induced by the uncertainty on the
mass-squared differences according to eq.~\eqref{eq:dmq}, and
$\sigma_{\rm obs}$ is the accuracy on $\Sum$ assumed for the
cosmological data. From eq.~\eqref{eq:min-values} we see that
$\sigma_{\rm osc}$ is below 1~meV for both orderings and $m_0 =
0$. For non-zero $m_0$, $\sigma_{\rm osc}$ is even smaller. Hence, for
$\sigma_{\rm obs} \gtrsim 0.01$~eV, the uncertainty on $\Sum$ from
oscillation data is negligible.

\begin{figure}[t]
  \centering \includegraphics[width=0.7\textwidth]{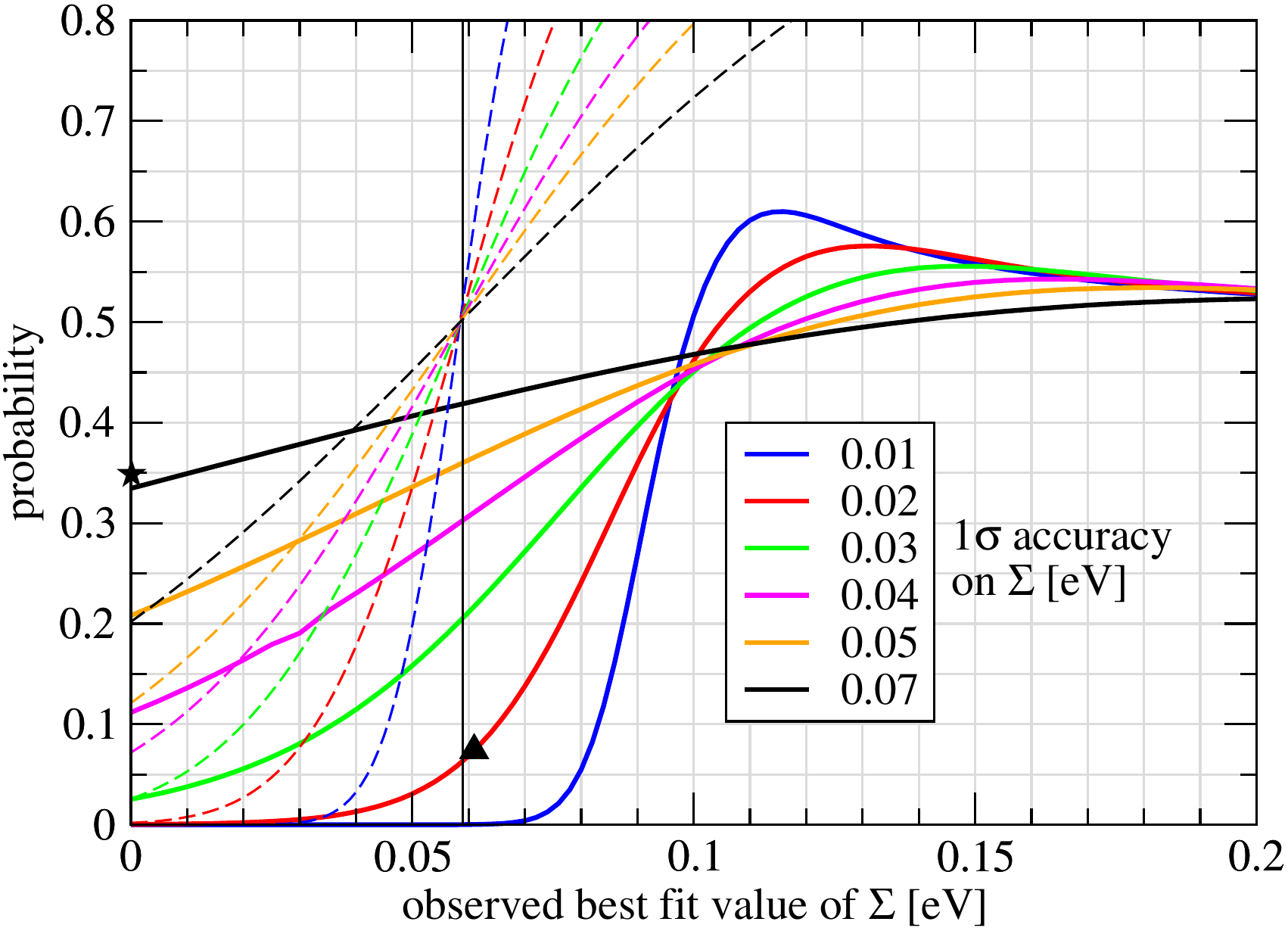}
  \mycaption{Illustration of the potential to exclude IO for a
    Gaussian toy likelihood. Solid curves show the probability of
    inverted ordering being correct as a function of the observed
    value of $\Sum$ for different assumptions about the obtained
    accuracy from cosmology, $\sigma_{\rm obs}$, according to the
    legend (values in eV). We assume equal prior probabilities for NO
    and IO. The dashed curves show the probability of observing a
    value of $\Sum$ equal or less than the one shown on the horizontal
    axis assuming that the true ordering is normal and $m_0 = 0$ for
    the assumed accuracy on $\Sum$. The thin vertical line indicates
    the median value for $\Sum$ for NO and $m_0=0$. The star and the
    triangle show approximately the cases of current and prospective
    data, respectively, as analysed in sec.~\ref{sec:data}.
    \label{fig:gauss}}
\end{figure}

The results based on this toy model for the likelihood are shown in
fig.~\ref{fig:gauss}. Solid curves in the plot show the probability of
IO, $p_I$, as a function of $\Sum^{\rm obs}$ for assumed values for
$\sigma_{\rm obs}$ ranging from 0.07~eV (corresponding approximately
to current data) down to 0.01~eV. Clearly, with an accuracy of order
$\gtrsim 0.05$~eV no meaningful statement can be made about the validity of
IO. We find that in order to reject IO with a confidence greater than
95\% (i.e., $p_I < 0.05$) accuracies of cosmological data of
$\sigma_{\rm obs} \lesssim 0.02$~eV are needed, in agreement with the
simple estimate provided in the introduction. If $\sigma_{\rm obs} =
0.03$~eV, the probability of observing a value of $\Sum$ such that
$p_I < 0.05$ is less than 10\%, if $m_0 = 0$ (for non-zero $m_0$ the
probability is smaller).

The star and the triangle in the plot indicate approximately the
cases corresponding to present data and EUCLID-like data,
respectively, as considered above. We observe that the Gaussian
toy-likelihood reproduces quite accurately the results for $p_I$
obtained for those two cases in section~\ref{sec:data}, justifying the
use of this model to estimate the required sensitivity.

\section{Conclusions}

If the neutrino mass ordering is normal and the spectrum is
hierarchical ($m_0 \ll \sqrt{\Delta m^2_{31}}$) cosmological data has
the potential to reject the hypothesis of inverted ordering by
constraining the sum of the neutrino masses sufficiently well. We
apply Bayes theorem to quantify possible evidence against inverted
ordering using present cosmological data as well as simulated data
from a future EUCLID-like mission. Our method provides a straight
forward way to combine cosmology with oscillation data by including a
possible preference for an ordering from oscillation data into the
prior probabilities for NO and IO.

For present cosmological data, we adopt a particular analysis of
Planck CMB + BAO data within the minimal $\Lambda$CDM model (6
parameters + \Sum), which leads to the constraint $\Sum < 0.14$~eV at
95\%~CL. For this analysis, our recipe gives posterior odds for normal
versus inverted ordering of about 2:1 (the posterior probability for
IO is 35\%).  Combining current cosmological and oscillation
data~\cite{Bergstrom:2015rba} we obtain posterior odds of about 3:2
(the posterior probability for IO is 39\%). As expected those results
show that current cosmological as well as oscillation data are not
sensitive to the mass ordering. However, this analysis provides an
example of how to quantify possible evidence against IO from
cosmological observation, and how to combine it with information from
oscillation data. Both of them are expected to become more sensitive
in the near future.

To illustrate this, we generate artificial data for a EUCLID-like
survey, assuming a fiducial model with NO and vanishing lightest
neutrino mass. Combined with CMB data, this data set would obtain an
accuracy to $\Sum$ of about 0.021 at $1\sigma$, sufficient to disfavour
IO at the 92\%~CL (corresponding to posterior odds of NO to IO of
about 12:1). Hence, we conclude that being able to exclude IO with
cosmology with significant confidence requires an accuracy on the sum
of neutrino masses of better than 0.02~eV.

We emphasize that statements about excluding the inverted ordering
with cosmology should be based on a proper statistical analysis. The
method we propose in section~\ref{sec:method} is based on Bayesian
statistics. Usually Bayesian methods are applied for the analysis of
cosmological data, and hence our method proposed here for the mass
ordering test fits consistently in this framework. Moreover, we can
deal in a straight forward way with the boundary implied by the
physics requirement that neutrino masses have to be non-negative.  Let
us emphasize, however, that the method proposed here is certainly not
unique, and it is possible to design alternative ways to report
cosmological information on the neutrino mass ordering. Hypothesis
tests based on frequentist statistics most likely will require
numerical simulations of the relevant test statistics because of the
physical boundary. In any case, we would like to encourage the
community to report possible evidence against the inverted mass
ordering using well defined statistical tools.

\subsection*{Acknowledgments}

This project has received funding from the European Union’s Horizon
2020 research and innovation programme under the Marie
Sklodowska-Curie grant agreement No 674896 (Elusives) as well as from
the Villum Foundation.

\bibliographystyle{JHEP_improved}
\bibliography{./refs}

\providecommand{\href}[2]{#2}\begingroup\raggedright\begin{thebibliography}{10}

\bibitem{Messier:2013sfa}
{\bf NOvA}, M.~D. Messier,
  \href{http://inspirehep.net/record/1246016/files/arXiv:1308.0106.pdf}{{\it
  {Extending the Nova Physics Program}}, } in {\em {Community Summer Study
  2013: Snowmass on the Mississippi (Cs$S^2$013) Minneapolis, MN, USA, July
  29-August 6, 2013}}, 2013.
\newblock \href{http://arxiv.org/abs/1308.0106}{{\tt 1308.0106}}.

\bibitem{Acciarri:2015uup}
{\bf DUNE}, R.~Acciarri et~al., {\it {Long-Baseline Neutrino Facility (LBNF)
  and Deep Underground Neutrino Experiment (DUNE) Conceptual Design Report
  Volume 2: the Physics Program for DUNE at LBNF}},
  \href{http://arxiv.org/abs/1512.06148}{{\tt 1512.06148}}.

\bibitem{Aartsen:2014oha}
{\bf IceCube PINGU}, M.~G. Aartsen et~al., {\it {Letter of Intent: the
  Precision Icecube Next Generation Upgrade (PINGU)}},
  \href{http://arxiv.org/abs/1401.2046}{{\tt 1401.2046}}.

\bibitem{Djurcic:2015vqa}
{\bf JUNO}, Z.~Djurcic et~al., {\it {JUNO Conceptual Design Report}},
  \href{http://arxiv.org/abs/1508.07166}{{\tt 1508.07166}}.

\bibitem{Ahmed:2015jtv}
{\bf ICAL}, S.~Ahmed et~al., {\it {Physics Potential of the Ical Detector at
  the India-Based Neutrino Observatory (INO)}},
  \href{http://arxiv.org/abs/1505.07380}{{\tt 1505.07380}}.

\bibitem{Blennow:2013oma}
M.~Blennow, P.~Coloma, P.~Huber, and T.~Schwetz,
  \href{http://dx.doi.org/10.1007/JHEP03(2014)028}{{\it {Quantifying the
  Sensitivity of Oscillation Experiments to the Neutrino Mass Ordering}}, }
  {\em JHEP} {\bf 03} (2014) 028, [\href{http://arxiv.org/abs/1311.1822}{{\tt
  1311.1822}}].

\bibitem{Lesgourgues:2004ps}
J.~Lesgourgues, S.~Pastor, and L.~Perotto,
  \href{http://dx.doi.org/10.1103/PhysRevD.70.045016}{{\it {Probing Neutrino
  Masses with Future Galaxy Redshift Surveys}}, } {\em Phys. Rev.} {\bf D70}
  (2004) 045016, [\href{http://arxiv.org/abs/hep-ph/0403296}{{\tt
  hep-ph/0403296}}].

\bibitem{Slosar:2006xb}
A.~Slosar, \href{http://dx.doi.org/10.1103/PhysRevD.73.123501}{{\it {Detecting
  Neutrino Mass Difference with Cosmology}}, } {\em Phys. Rev.} {\bf D73}
  (2006) 123501, [\href{http://arxiv.org/abs/astro-ph/0602133}{{\tt
  astro-ph/0602133}}].

\bibitem{Jimenez:2010ev}
R.~Jimenez, T.~Kitching, C.~Pena-Garay, and L.~Verde,
  \href{http://dx.doi.org/10.1088/1475-7516/2010/05/035}{{\it {Can We Measure
  the Neutrino Mass Hierarchy in the Sky?}}, } {\em JCAP} {\bf 1005} (2010)
  035, [\href{http://arxiv.org/abs/1003.5918}{{\tt 1003.5918}}].

\bibitem{Hamann:2012fe}
J.~Hamann, S.~Hannestad, and Y.~Y.~Y. Wong,
  \href{http://dx.doi.org/10.1088/1475-7516/2012/11/052}{{\it {Measuring
  Neutrino Masses with a Future Galaxy Survey}}, } {\em JCAP} {\bf 1211} (2012)
  052, [\href{http://arxiv.org/abs/1209.1043}{{\tt 1209.1043}}].

\bibitem{Gonzalez-Garcia:2014bfa}
M.~C. Gonzalez-Garcia, M.~Maltoni, and T.~Schwetz,
  \href{http://dx.doi.org/10.1007/JHEP11(2014)052}{{\it {Updated Fit to Three
  Neutrino Mixing: Status of Leptonic CP Violation}}, } {\em JHEP} {\bf 11}
  (2014) 052, [\href{http://arxiv.org/abs/1409.5439}{{\tt 1409.5439}}]. NuFit
  version~2.1 {\tt http://www.nu-fit.org}.

\bibitem{Ade:2015xua}
{\bf Planck}, P.~A.~R. Ade et~al., {\it {Planck 2015 Results. XIII.
  Cosmological Parameters}},  \href{http://arxiv.org/abs/1502.01589}{{\tt
  1502.01589}}.

\bibitem{Palanque-Delabrouille:2015pga}
N.~Palanque-Delabrouille et~al.,
  \href{http://dx.doi.org/10.1088/1475-7516/2015/11/011}{{\it {Neutrino masses
  and cosmology with Lyman-alpha forest power spectrum}}, } {\em JCAP} {\bf
  1511} (2015), no.~11 011, [\href{http://arxiv.org/abs/1506.05976}{{\tt
  1506.05976}}].

\bibitem{Cuesta:2015iho}
A.~J. Cuesta, V.~Niro, and L.~Verde,
  \href{http://dx.doi.org/10.1016/j.dark.2016.04.005}{{\it {Neutrino Mass
  Limits: Robust Information from the Power Spectrum of Galaxy Surveys}}, }
  {\em Phys. Dark Univ.} {\bf 13} (2016) 77--86,
  [\href{http://arxiv.org/abs/1511.05983}{{\tt 1511.05983}}].

\bibitem{Huang:2015wrx}
Q.-G. Huang, K.~Wang, and S.~Wang, {\it {Constraints on the Neutrino Mass and
  Mass Hierarchy from Cosmological Observations}},
  \href{http://arxiv.org/abs/1512.05899}{{\tt 1512.05899}}.

\bibitem{DiValentino:2015sam}
E.~Di~Valentino, E.~Giusarma, O.~Mena, A.~Melchiorri, and J.~Silk,
  \href{http://dx.doi.org/10.1103/PhysRevD.93.083527}{{\it {Cosmological Limits
  on Neutrino Unknowns Versus Low Redshift Priors}}, } {\em Phys. Rev.} {\bf
  D93} (2016), no.~8 083527, [\href{http://arxiv.org/abs/1511.00975}{{\tt
  1511.00975}}].

\bibitem{Giusarma:2016phn}
E.~Giusarma, M.~Gerbino, O.~Mena, S.~Vagnozzi, S.~Ho, et~al., {\it {On the
  Improvement of Cosmological Neutrino Mass Bounds}},
  \href{http://arxiv.org/abs/1605.04320}{{\tt 1605.04320}}.

\bibitem{Blennow:2013kga}
M.~Blennow, \href{http://dx.doi.org/10.1007/JHEP01(2014)139}{{\it {On the
  Bayesian Approach to Neutrino Mass Ordering}}, } {\em JHEP} {\bf 01} (2014)
  139, [\href{http://arxiv.org/abs/1311.3183}{{\tt 1311.3183}}].

\bibitem{Hall:2012kg}
A.~C. Hall and A.~Challinor,
  \href{http://dx.doi.org/10.1111/j.1365-2966.2012.21493.x}{{\it {Probing the
  Neutrino Mass Hierarchy with CMB Weak Lensing}}, } {\em Mon. Not. Roy.
  Astron. Soc.} {\bf 425} (2012) 1170--1184,
  [\href{http://arxiv.org/abs/1205.6172}{{\tt 1205.6172}}].

\bibitem{Bergstrom:2015rba}
J.~Bergstrom, M.~C. Gonzalez-Garcia, M.~Maltoni, and T.~Schwetz,
  \href{http://dx.doi.org/10.1007/JHEP09(2015)200}{{\it {Bayesian Global
  Analysis of Neutrino Oscillation Data}}, } {\em JHEP} {\bf 09} (2015) 200,
  [\href{http://arxiv.org/abs/1507.04366}{{\tt 1507.04366}}].

\bibitem{DiValentino:2016hlg}
E.~Di~Valentino, A.~Melchiorri, and J.~Silk, {\it {Reconciling Planck with the
  local value of $H_0$ in extended parameter space}},
  \href{http://arxiv.org/abs/1606.00634}{{\tt 1606.00634}}.

\bibitem{Lu:2016hsd}
J.~Lu, M.~Liu, Y.~Wu, Y.~Wang, and W.~Yang, {\it {Cosmic Constraint on Massive
  Neutrinos in Viable F(R) Gravity with Producing Lcdm Background Expansion}},
  \href{http://arxiv.org/abs/1606.02987}{{\tt 1606.02987}}.

\bibitem{Canac:2016smv}
N.~Canac, G.~Aslanyan, K.~N. Abazajian, R.~Easther, and L.~C. Price, {\it
  {Testing for New Physics: Neutrinos and the Primordial Power Spectrum}},
  \href{http://arxiv.org/abs/1606.03057}{{\tt 1606.03057}}.

\bibitem{Beutler:2011hx}
F.~Beutler, C.~Blake, M.~Colless, D.~H. Jones, L.~Staveley-Smith, et~al.,
  \href{http://dx.doi.org/10.1111/j.1365-2966.2011.19250.x}{{\it {The 6dF
  Galaxy Survey: Baryon Acoustic Oscillations and the Local Hubble Constant}},
  } {\em Mon. Not. Roy. Astron. Soc.} {\bf 416} (2011) 3017--3032,
  [\href{http://arxiv.org/abs/1106.3366}{{\tt 1106.3366}}].

\bibitem{Ross:2014qpa}
A.~J. Ross, L.~Samushia, C.~Howlett, W.~J. Percival, A.~Burden, et~al.,
  \href{http://dx.doi.org/10.1093/mnras/stv154}{{\it {The clustering of the
  SDSS DR7 main Galaxy sample – I. A 4 per cent distance measure at $z =
  0.15$}}, } {\em Mon. Not. Roy. Astron. Soc.} {\bf 449} (2015), no.~1
  835--847, [\href{http://arxiv.org/abs/1409.3242}{{\tt 1409.3242}}].

\bibitem{Anderson:2012sa}
L.~Anderson et~al.,
  \href{http://dx.doi.org/10.1111/j.1365-2966.2012.22066.x}{{\it {The
  clustering of galaxies in the SDSS-III Baryon Oscillation Spectroscopic
  Survey: Baryon Acoustic Oscillations in the Data Release 9 Spectroscopic
  Galaxy Sample}}, } {\em Mon. Not. Roy. Astron. Soc.} {\bf 427} (2013), no.~4
  3435--3467, [\href{http://arxiv.org/abs/1203.6594}{{\tt 1203.6594}}].

\bibitem{Anderson:2013zyy}
{\bf BOSS}, L.~Anderson et~al.,
  \href{http://dx.doi.org/10.1093/mnras/stu523}{{\it {The clustering of
  galaxies in the SDSS-III Baryon Oscillation Spectroscopic Survey: baryon
  acoustic oscillations in the Data Releases 10 and 11 Galaxy samples}}, } {\em
  Mon. Not. Roy. Astron. Soc.} {\bf 441} (2014), no.~1 24--62,
  [\href{http://arxiv.org/abs/1312.4877}{{\tt 1312.4877}}].

\bibitem{Riess:2016jrr}
A.~G. Riess et~al., {\it {A 2.4\% Determination of the Local Value of the
  Hubble Constant}},  \href{http://arxiv.org/abs/1604.01424}{{\tt 1604.01424}}.

\bibitem{Lewis:2002ah}
A.~Lewis and S.~Bridle,
  \href{http://dx.doi.org/10.1103/PhysRevD.66.103511}{{\it {Cosmological
  parameters from CMB and other data: A Monte Carlo approach}}, } {\em Phys.
  Rev.} {\bf D66} (2002) 103511,
  [\href{http://arxiv.org/abs/astro-ph/0205436}{{\tt astro-ph/0205436}}].

\bibitem{LSST}
 Large Synoptic Survey Telescope (LSST) {\tt http://www.lsst.org}.

\bibitem{euclid}
 Euclid ESA mission {\tt http://sci.esa.int/euclid}.

\bibitem{Audren:2012vy}
B.~Audren, J.~Lesgourgues, S.~Bird, M.~G. Haehnelt, and M.~Viel,
  \href{http://dx.doi.org/10.1088/1475-7516/2013/01/026}{{\it {Neutrino masses
  and cosmological parameters from a Euclid-like survey: Markov Chain Monte
  Carlo forecasts including theoretical errors}}, } {\em JCAP} {\bf 1301}
  (2013) 026, [\href{http://arxiv.org/abs/1210.2194}{{\tt 1210.2194}}].

\bibitem{Basse:2013zua}
T.~Basse, O.~E. Bjaelde, J.~Hamann, S.~Hannestad, and Y.~Y.~Y. Wong,
  \href{http://dx.doi.org/10.1088/1475-7516/2014/05/021}{{\it {Dark energy
  properties from large future galaxy surveys}}, } {\em JCAP} {\bf 1405} (2014)
  021, [\href{http://arxiv.org/abs/1304.2321}{{\tt 1304.2321}}].

\bibitem{Cerbolini:2013uya}
M.~C.~A. Cerbolini, B.~Sartoris, J.-Q. Xia, A.~Biviano, S.~Borgani, et~al.,
  \href{http://dx.doi.org/10.1088/1475-7516/2013/06/020}{{\it {Constraining
  neutrino properties with a Euclid-like galaxy cluster survey}}, } {\em JCAP}
  {\bf 1306} (2013) 020, [\href{http://arxiv.org/abs/1303.4550}{{\tt
  1303.4550}}].

\bibitem{Font-Ribera:2013rwa}
A.~Font-Ribera, P.~McDonald, N.~Mostek, B.~A. Reid, H.-J. Seo, et~al.,
  \href{http://dx.doi.org/10.1088/1475-7516/2014/05/023}{{\it {DESI and other
  dark energy experiments in the era of neutrino mass measurements}}, } {\em
  JCAP} {\bf 1405} (2014) 023, [\href{http://arxiv.org/abs/1308.4164}{{\tt
  1308.4164}}].

\bibitem{Abazajian:2013oma}
K.~N. Abazajian et~al.,
  \href{http://dx.doi.org/10.1016/j.astropartphys.2014.05.014}{{\it {Neutrino
  Physics from the Cosmic Microwave Background and Large Scale Structure}}, }
  {\em Astropart. Phys.} {\bf 63} (2015) 66--80,
  [\href{http://arxiv.org/abs/1309.5383}{{\tt 1309.5383}}].

\bibitem{Allison:2015qca}
R.~Allison, P.~Caucal, E.~Calabrese, J.~Dunkley, and T.~Louis,
  \href{http://dx.doi.org/10.1103/PhysRevD.92.123535}{{\it {Towards a
  Cosmological Neutrino Mass Detection}}, } {\em Phys. Rev.} {\bf D92} (2015),
  no.~12 123535, [\href{http://arxiv.org/abs/1509.07471}{{\tt 1509.07471}}].

\bibitem{Basse:2014qqa}
T.~Basse, J.~Hamann, S.~Hannestad, and Y.~Y.~Y. Wong,
  \href{http://dx.doi.org/10.1088/1475-7516/2015/06/042}{{\it {Getting leverage
  on inflation with a large photometric redshift survey}}, } {\em JCAP} {\bf
  1506} (2015), no.~06 042, [\href{http://arxiv.org/abs/1409.3469}{{\tt
  1409.3469}}].

\bibitem{Jimenez:2016ckl}
R.~Jimenez, C.~Pena-Garay, and L.~Verde, {\it {Neutrino footprint in Large
  Scale Structure}},  \href{http://arxiv.org/abs/1602.08430}{{\tt 1602.08430}}.

\end{thebibliography}\endgroup

\end{document}